\documentclass[reprint,amsmath,amssymb,aps,pre,showpacs,eqsecnum]{revtex4-1}

\usepackage{natbib}
\usepackage{amssymb,amsmath}

\usepackage[linktoc=all,unicode=true,pdfusetitle,  bookmarks=true,bookmarksnumbered=false,bookmarksopen=false, breaklinks=true,pdfborder={0 0 0},backref=false,colorlinks=true]{hyperref}
\hypersetup{linkcolor=blue,citecolor=red}

\usepackage[hyphenbreaks]{breakurl}

\usepackage{epsfig}
\usepackage{bm}
\usepackage{color}
\usepackage{graphicx}

\newcommand\beq{\begin{equation}}
\newcommand\eeq{\end{equation}}
\newcommand\beqa{\begin{eqnarray}}
\newcommand\eeqa{\end{eqnarray}}
\newcommand{\dd}{\text{d}}
\newcommand{\nn}{\nonumber\\}
\def\bal#1\eal{\begin{align}#1\end{align}}

\newcommand{\di}{{(d)}}
\newcommand{\con}{{\text{cont}}}
\newcommand{\vir}{{\text{vir}}}
\newcommand{\comp}{{\text{comp}}}
\newcommand{\eeta}{\phi}
\newcommand{\on}{{(1)}}
\newcommand{\thr}{{(3)}}

\begin{document}
\title{Radial distribution function for hard spheres in fractal dimensions: A heuristic approximation}

\author{Andr\'es Santos}
\email{andres@unex.es}
\homepage{http://www.unex.es/eweb/fisteor/andres/Cvitae/}
\affiliation{Departamento de F\'{\i}sica and Instituto de Computaci\'on Cient\'ifica Avanzada (ICCAEx), Universidad de
Extremadura, Badajoz, E-06071, Spain}
\author{Mariano L\'opez de Haro}
\email{malopez@unam.mx}
\homepage{http://xml.ier.unam.mx/xml/tc/ft/mlh/}
\affiliation{Instituto de Energ\'{\i}as Renovables, Universidad Nacional Aut\'onoma de M\'exico (U.N.A.M.), Temixco, Morelos 62580, M{e}xico}
\date{\today}
\begin{abstract}
Analytic approximations for the radial distribution function, the structure factor, and the equation of state of hard-core  fluids in fractal dimension $d$ ($1 \leq d \leq 3$) are developed as heuristic interpolations from the knowledge of the exact and Percus--Yevick results for the hard-rod  and hard-sphere fluids, respectively. In order to assess their value, such approximate results are compared with those of recent Monte Carlo simulations and numerical solutions of the Percus--Yevick equation for a fractal dimension [M. Heinen \emph{et al.}, Phys.\ Rev.\ Lett.\ \textbf{115}, 097801 (2015)], a good agreement being observed.
\end{abstract}
\maketitle

\maketitle
\section{Introduction}
The so-called classical fluids can, to a good approximation, be modeled and treated by the methods of classical statistical mechanics. Among these systems, the relatively simple hard-core models (hard rods, disks, spheres, and hyperspheres) have played a very important role in laying the foundations of a solid theoretical framework for dealing with the thermodynamic and structural properties of real fluids. Among their assets, it should be pointed out that all hard-core fluids share similar properties, including the existence of a fluid-solid phase transition (except in the one-dimensional case) that occurs at a packing fraction that becomes smaller as the dimensionality $d$ is increased. Thus, increasing the dimensionality of these hard-core fluids leads in general to simpler mathematical formulations to describe similar phenomenology \cite{CIPZ2011}. On the other hand, while real fluids in the bulk are three-dimensional, once they are confined their \emph{effective} dimension becomes $d=2$ or $d=1$. It is well known that confinement has a substantial influence on the thermodynamic and structural properties of fluids, a subject that has been profusely dealt with in the literature (see, for instance, Refs.\ \onlinecite{L09,FLS12,FLS13} and references therein).

On the other hand, it is only very recently that Heinen \emph{et al.}\ \cite{HSBL15} have addressed the problem of formulating a theory for the fractal analog of the simplest classical generic model fluid in integer dimensions, namely, the hard-core fluid in fractal dimensions between $d=1$ and $3$. In their model, they consider fractal particles in a fractal configuration space, both of the same noninteger dimension.
Such a generic model of \emph{fractal liquids} can describe, for instance, microphase separated binary liquids in porous media and highly branched liquid droplets confined to a fractal polymer backbone in a gel.
Heinen \emph{et al.}\ \cite{HSBL15} performed Monte Carlo (MC) simulations of  fractal hard ``spheres'' lying on a near-critical percolating lattice cluster. In this context, a fractal sphere of ``diameter'' $\sigma$ is defined by those lattice sites at a \emph{chemical distance} (taxicab metric) from the sphere center site smaller than $\frac{1}{2}\sigma$. Moreover,  Heinen \emph{et al.}\ were able to (numerically) generalize  the solution to the Ornstein--Zernike equation with the Percus--Yevick (PY) closure to  noninteger dimensions. In that solution, the corresponding radial distribution function (RDF) and thermodynamic properties represent the analytic continuations of the standard PY theory with respect to the dimension.

The aim of this paper is to develop an analytic approximation for the RDF $g^\di(r,\eeta)$ of hard-core fluids in fractal dimension $d$ (with $1 \leq d \leq 3$), where  $r$ is the distance (in the  non-Euclidian fractal space) and  $\eeta\equiv [(\pi/4)^{d/2}/\Gamma(1+d/2)]\rho\sigma^d$ is the packing fraction (with $\sigma$ the hard-core diameter and $\rho$ the number density).
Following a heuristic approach similar to the one used  to derive the RDF of a hard-disk fluid ($d=2$) \cite{YS93c}, our approximation will be constructed simply as an interpolation between the exact RDF $g^{(1)}(r,\eeta)$ at $d=1$ \cite{HGM34,T36,HC04,S14,S16,LNP_book_note_15_06_1} and the PY solution $g^{(3)}(r,\eeta)$ at $d=3$ \cite{T63,W63,W64,HM06,S14,S16,LNP_book_note_13_10}, with suitably scaled packing fractions. The determination of the two scaling parameters requires the independent proposal of an analytic approximate expression for the contact value $g_\con^\di(\eeta)\equiv g^\di(\sigma^+,\eeta)$, and hence  for the virial equation of state of the fractal fluid, and this represents an extra bonus of our heuristic approach. As we will see, despite the simplicity of the theory, the results agree fairly well with both MC simulations and PY numerical results obtained in Ref.\ \cite{HSBL15} at $d=1.67659$.

The paper is organized as follows. In Sec.\ \ref{sec2}, we propose the explicit expressions for $g_\con^\di(\eeta)$ and $g^\di(r,\eeta)$, and derive the compressibility factor $Z^\di(\eeta)$ that follows from both the virial and compressibility routes to the equation of state. In Sec.\ \ref{Results1}, we perform an analysis of the behavior of the third and fourth virial coefficients that follow from our formulation, while  our theoretical results for the compressibility factor, the RDF, and the structure factor are compared in Sec.\ \ref{Results2} with those of the generalized PY theory and the MC data presented in Ref.\ \onlinecite{HSBL15}. The paper concludes in Sec.\ \ref{concl} with some concluding remarks.

\section{Radial distribution function and equation of state}
\label{sec2}
Due to their importance in our later development, before constructing the function $g^\di(r,\eeta)$, let us first consider two different quantities (one local and another one global) related to $g^\di(r,\eeta)$. On the one hand, we take the contact value $g^\di_{\con}(\eeta)$ defined above.
The exact and PY expressions for this quantity in the cases $d=1$ and $d=3$ are, respectively,
\begin{subequations}
\label{g1-1D&3D}
\beq
g^{(1)}_{\con}(\eeta)=\frac{1}{1-\eeta},
\label{g1-1D}
\eeq
\beq
g^{(3)}_{\con}(\eeta)=\frac{1+\eeta/2}{(1-\eeta)^2}.
\label{g1-3D}
\eeq
\end{subequations}
We also note that the scaled-particle theory (SPT) approximation \cite{RFL59,HFL61} for $d=2$ has the form
\beq
g^{(2)}_\con(\eeta)=\frac{1-a^{(2)}\eeta}{(1-\eeta)^2},
\label{g1-2D}
\eeq
with $a^{(2)}=\frac{1}{2}$. A more accurate  fractional value  $a^{(2)}=\frac{7}{16}=0.4375$ was proposed by Henderson \cite{H75}, who also noticed that the value $a^{(2)}=2\sqrt{3}/\pi-2/3\simeq 0.436$ guarantees that Eq.\ \eqref{g1-2D} reproduces the exact third virial coefficient of hard disks. Here we will adopt the latter value for $a^{(2)}$.

In view of Eqs.\ \eqref{g1-1D&3D} and \eqref{g1-2D}, it is suggestive to construct a simple generalization  for $1\leq d\leq 3$ as
\begin{subequations}
\label{g1-dD}
\beq
g^{(d)}_\con(\eeta)=\frac{1-a^\di\eeta}{(1-\eeta)^2}.
\eeq
The density-independent  coefficient $a^\di$ can be constructed as a quadratic polynomial in $d$ with coefficients such that $a^{(1)}=1$, $a^{(2)}=2\sqrt{3}/\pi-2/3$, and $a^{(3)}=-\frac{1}{2}$. The resulting expression is
\beq
a^\di=\frac{1}{4}(5-d)(2-d)+(3-d)(d-1)a^{(2)}.
\label{g1-dD_a2}
\eeq
\end{subequations}

Next, we consider a convenient \emph{global} quantity related to $g^\di(r,\eeta)$, namely the first moment of the total correlation function, $h^\di(r,\eeta)=g^\di(r,\eeta)-1$, defined as
\beq
H^\di(\eeta)=-\sigma^{-2}\int_0^\infty \dd r\, r h^\di(r,\eeta).
\label{1}
\eeq
The exact and PY results for this quantity corresponding to $d=1$ and $d=3$ are \cite{YS93c}, respectively,
\begin{subequations}
\label{2+3}
\beq
H^{(1)}(\eeta)=\frac{1}{2}-\frac{2}{3}\eeta+\frac{1}{4}\eeta^2,
\label{2}
\eeq
\beq
H^{(3)}(\eeta)=\frac{\frac{1}{2}-\frac{1}{20}\eeta(2-\eeta)}{1+2\eeta}.
\label{3}
\eeq
\end{subequations}

Note that the knowledge of $g^\di_\con(\eeta)$ determines the equation of state via the virial route as \cite{HM06,S16}
\beq
Z_\vir^\di(\eeta)=1+2^{d-1}\eeta g^\di_\con(\eeta),
\label{Z}
\eeq
where $Z^\di\equiv p/\rho k_BT$ is the so-called compressibility factor ($p$, $k_B$, and $T$ being the pressure, Boltzmann constant, and absolute temperature, respectively).
On the other hand, the compressibility equation of state may be derived from the following relations \cite{HM06,S16}:
\begin{subequations}
\label{chi_d}
\bal
\chi^\di(\eeta)\equiv&\left[\frac{\partial \eeta Z^\di(\eeta)}{\partial \eeta}\right]^{-1}=1+\rho\int\dd\mathbf{r}\,h^\di(r,\eeta)\nn
=&1+d2^{d}\eeta\sigma^{-d}\int_0^\infty \dd r\, r^{d-1}h^\di(r,\eeta),
\label{chi_dd}
\eal
\beq
Z_\comp^\di(\eeta)=\int_0^1\frac{\dd t}{\chi^\di(\eeta t)}.
\label{Zcomp}
\eeq
\end{subequations}
In particular, in the two-dimensional case ($d=2$) the moment $H^{(2)}$ is directly related to the isothermal compressibility:
\beq
\chi^{(2)}(\eeta)=1-8\eeta H^{(2)}(\eeta).
\label{chi_2}
\eeq
 Imposing thermodynamic consistency with Eqs.\ \eqref{g1-2D}, \eqref{Z}, and \eqref{chi_2}, one obtains
\beq
H^{(2)}(\eeta)=\frac{\frac{1}{2}-\frac{1}{4}a^{(2)}\eeta(3-\eeta)}{1+\eeta+[1-2a^{(2)}]\eeta^2(3-\eeta)}.
\label{4}
\eeq
The simplest common structure of Eqs.\ \eqref{2+3} and \eqref{4} can be seen to consist in the ratio between a quadratic and a cubic function of $\eeta$.
Analogously to what we did in the case of the generalization \eqref{g1-dD}, we keep such a structure for all $1\leq d\leq 3$ with density-independent coefficients expressed as quadratic functions of $d$. The result is
\begin{subequations}
\label{5}
\beq
H^\di(\eeta)=\frac{\frac{1}{2}-A^\di\eeta+C^\di\eeta^2}{1+(d-1)\eeta\left\{1+(3-d)[1-2a^{(2)}]\eeta(3-\eeta)\right\}},
\eeq
with
\beq
A^\di=\frac{1}{60}(2-d)(63-23d)+\frac{3}{4}(d-1)(3-d)a^{(2)},
\eeq
\beq
C^\di=\frac{1}{20}(2-d)(8-3d)+\frac{1}{4}(d-1)(3-d)a^{(2)}.
\eeq
\end{subequations}

Now we turn to the construction of a function $g^\di(r,\eeta)$ that, apart from reducing to $g^{(1)}(r,\eeta)$ and $g^{(3)}(r,\eeta)$ in the limits $d\to 1$ and $d\to 3$, respectively, is consistent with Eqs.\ \eqref{g1-dD} and \eqref{5}. Following the heuristic idea of Ref.\ \onlinecite{YS93c}, we assume the simple interpolation formula
\bal
g^\di(r,\eeta)=&\alpha^\di(\eeta) g^{(1)}\left(r,\lambda_1^\di(\eeta)\eeta\right)\nn
&+[1-\alpha^\di(\eeta)]g^{(3)}\left(r,\lambda_3^\di(\eeta)\eeta\right),
\label{6}
\eal
where the mixing parameter $\alpha^\di(\eeta)$ and the scaling factors $\lambda_1^\di(\eeta)$ and $\lambda_3^\di(\eeta)$ are functions of $\eeta$ and $d$ to be determined.
The exact and PY expressions for $g^\on(r,\eeta)$ and $g^\thr(r,\eeta)$, respectively, are recalled in the Appendix.

To fix $\lambda_1^\di(\eeta)$ and $\lambda_3^\di(\eeta)$, and again as in Ref.\ \onlinecite{YS93c},
we impose consistency with \eqref{g1-dD} \emph{regardless of}  the choice of the mixing parameter $\alpha^\di$. This implies that
\beq
g^\di_\con(\eeta)=g^{(1)}_\con\left(\lambda_1^\di(\eeta)\eeta\right)=g^{(3)}_\con\left(\lambda_3^\di(\eeta)\eeta\right).
\label{7}
\eeq
The scaling factors $\lambda_1^\di(\eeta)$ and  $\lambda_3^\di(\eeta)$ have the following interpretation. Given a $d$-dimensional system at a packing fraction $\eeta_d$, the \emph{effective} packing fractions $\eeta_1$ and $\eeta_3$ of the \emph{reference} one- and three-dimensional systems (i.e., those having the same contact value), are, respectively,
\begin{subequations}
\label{2.6}
  \bal
  \eeta_1=&\lambda_1^\di(\eeta_d)\eeta_d,
  \label{2.6a}\\
  \eeta_3=&\lambda_3^\di(\eeta_d)\eeta_d.
  \label{2.6b}
  \eal
\end{subequations}
In particular, setting $d=1$ in Eq.\ \eqref{2.6b} yields $\eeta_3=\lambda_3^{(1)}(\eeta_1)\eeta_1$. Combining this with Eq.\ \eqref{2.6b}, we obtain
$\lambda_3^{(1)}(\eeta_1)\eeta_1=\lambda_3^\di(\eeta_d)\eeta_d$.
Next, taking into account Eq.\ \eqref{2.6a} we derive the following consistency condition
\begin{subequations}
\label{2.8}
\beq
\lambda_3^{(1)}\left(\lambda_1^\di(\eeta_d)\eeta_d\right)=\frac{\lambda_3^\di(\eeta_d)}{\lambda_1^\di(\eeta_d)}.
\label{2.8a}
\eeq
Proceeding in a similar way, we also have
\beq
\lambda_1^{(3)}\left(\lambda_3^\di(\eeta_d)\eeta_d\right)=\frac{\lambda_1^\di(\eeta_d)}{\lambda_3^\di(\eeta_d)}.
\label{2.8b}
\eeq
\end{subequations}
It can be easily checked that Eqs.\ \eqref{2.8} are indeed satisfied as a consequence of Eq.\ \eqref{7}.

\begin{figure}[h]
 \includegraphics[width=0.95\columnwidth]{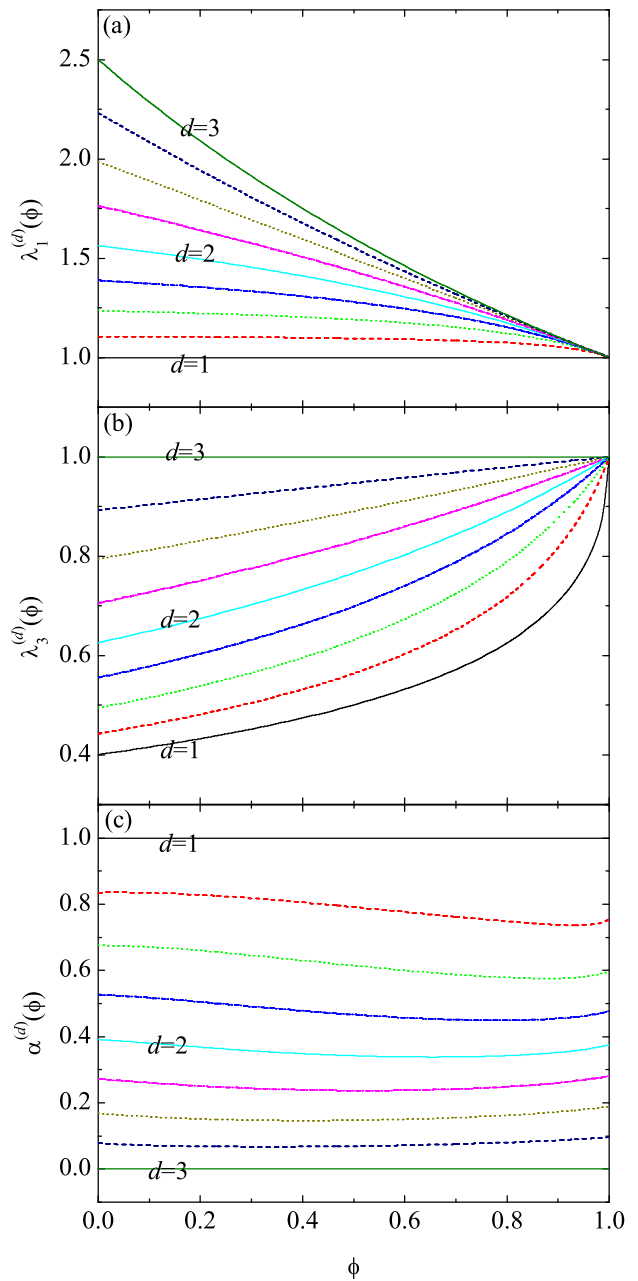}
\caption{Plot of (a) the scaling factor $\lambda_1^\di(\eeta)$, (b) the scaling factor $\lambda_3^\di(\eeta)$, and (c) the mixing parameter $\alpha^\di(\eeta)$ as functions of $\eeta$ for  $d=1, 1.25,1.5, 1.75,2,2.25,2.5,2.75,3$.}\label{fig:lambda_alpha}
\end{figure}

Using Eqs.\ \eqref{g1-1D&3D}, the solution to Eq.\ \eqref{7} is easily found to be
\beq
\label{8}
\lambda_1^\di(\eeta)=\frac{g^\di_\con(\eeta)-1}{\eeta g^\di_\con(\eeta)},
\eeq
\beq
\label{9}
\lambda_3^\di(\eeta)=\frac{1+4g^\di_\con(\eeta)-\sqrt{1+24g^\di_\con(\eeta)}}{4\eeta g^\di_\con(\eeta)}.
\eeq
Once the scaling parameters $\lambda_1^\di(\eeta)$ and $\lambda_3^\di(\eeta)$ are determined, we obtain the mixing parameter $\alpha^\di(\eeta)$ by imposing consistency between Eq.\ \eqref{6} and the moment \eqref{5}:
\beq
\alpha^\di(\eeta)=\frac{H^\di(\eeta)-H^{(3)}\left(\lambda_3^\di(\eeta)\eeta\right)}{H^{(1)}\left(\lambda_1^\di(\eeta)\eeta\right)-H^{(3)}\left(\lambda_3^\di(\eeta)\eeta\right)}.
\label{10}
\eeq

In summary, our proposal is defined by Eqs.\ \eqref{6} and \eqref{8}--\eqref{10}, with $g_\con^\di(\eeta)$ and $H^\di(\eeta)$ being given by Eqs.\ \eqref{g1-dD} and \eqref{5}, respectively.
In turn,  the compressibility factor $Z^\di(\eeta)$ can be obtained either analytically from the virial route \eqref{Z}, with the contact value $g_\con^\di(\eeta)$ given  by Eqs.\ \eqref{g1-dD}, or numerically from the compressibility route \eqref{chi_d}.
By construction, the approximation \eqref{6} reduces to the exact and PY results in the limits $d\to 1$ and $d\to 3$, respectively, and is consistent  (via both the virial and compressibility routes) with Henderson's equation of state [see Eqs.\ \eqref{g1-2D} and \eqref{4}] in the limit $d\to 2$.

Note that the same method could still be applied by prescribing for $g^\thr(r,\eeta)$ a RDF thermodynamically consistent and more accurate than the PY one, such as the rational-function approximation \cite{YS91,YHS96,HYS08,S16}, and/or by enforcing an equation of state for hard-disk fluids different from Henderson's. For simplicity, however, we keep the approximation \eqref{6} as formulated above \cite{note_16_04_1}.

Figure \ref{fig:lambda_alpha} shows $\lambda_1^\di(\eeta)$, $\lambda_3^\di(\eeta)$, and $\alpha^\di(\eeta)$ as functions of $\eeta$ for the representative values  $d=1,1.25,\ldots, 2, 2.25,\ldots,3$. As expected on physical grounds, $\lambda_1^\di(\eeta)$ and $\lambda_3^\di(\eeta)$ are, respectively, monotonically decreasing and increasing functions of $\eeta$. On the other hand, $\alpha^\di(\eeta)$ is very weakly dependent on $\eeta$. Obviously, $\lambda_1^{(1)}(\eeta)=1$, $\lambda_3^{(3)}(\eeta)=1$, $\alpha^{(1)}(\eeta)=1$, and $\alpha^{(3)}(\eeta)=0$.

Before closing this section, let us also consider another structural property, namely the structure factor
\beq
S^\di(k,\eeta)=1+\rho \widetilde{h}^\di(k,\eeta).
\label{S_k}
\eeq
Here, the Fourier transform of the total correlation function is \cite{HSBL15}
\bal
\widetilde{h}^\di(k,\eeta)=&\int \dd \mathbf{r}\, e^{i\mathbf{k}\cdot\mathbf{r}} h^\di(r,\eeta)\nn
=&\frac{(2\pi)^{d/2}}{k^{d/2-1}}\int_0^\infty\dd r\, r^{d/2}h^\di(r,\eeta)J_{d/2-1}(kr),
\label{h_k}
\eal
where $J_\nu(x)$ is the Bessel function. Insertion of Eq.\ \eqref{6} yields
\bal
S^\di(k,\eeta)=&1+\frac{\alpha^\di(\eeta)}{\lambda_1^\di(\eeta)}I_1^\di\left(k,\lambda_1^\di(\eeta)\eeta\right)\nn
&+\frac{1-\alpha^\di(\eeta)}{\lambda_3^\di(\eeta)}I_3^\di\left(k,\lambda_3^\di(\eeta)\eeta\right),
\label{S_kk}
\eal
where we have called
\bal
I_{1,3}^\di(k,\eeta)\equiv& \frac{2^{3d/2}\Gamma(1+d/2)}{k^{d/2-1}}\sigma^{-d}\eeta\nn
&\times\int_0^\infty\dd r\, r^{d/2}h^{(1,3)}(r,\eeta)J_{d/2-1}(kr).
\label{I_k}
\eal

\section{Low-density properties}
\label{Results1}

\begin{figure}
 \includegraphics[width=0.95\columnwidth]{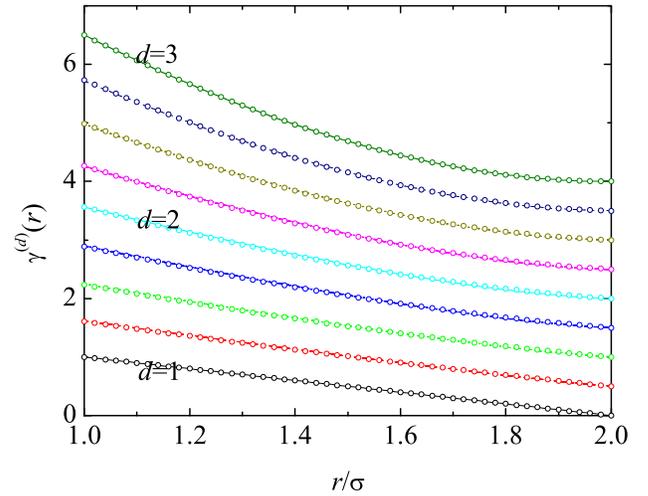}
\caption{Function $\gamma^\di(r)$ for  $d=1, 1.25,1.5, 1.75,2,2.25,2.5,2.75,3$. The curves and the circles correspond to the exact function \eqref{12} and the approximation \eqref{15}, respectively. Note that, for better visibility, the curves have been shifted vertically a distance $0, 0.5,\ldots, 4$ for $d=1,1.25,\ldots,3$, respectively.}\label{fig:gamma}
\end{figure}

In order to illustrate some of the consequences of our proposal, in this section we analyze its low-density predictions.
To first order in density, one can write
\beq
g^\di(r,\eeta)=\Theta(r-1)\left[1+\gamma^\di(r)\eeta+\mathcal{O}(\eeta^2)\right],
\label{11}
\eeq
where, without loss of generality, henceforth we have set the value of the hard-core diameter to be $\sigma=1$ (i.e., distances are measured in units of the hard-core diameter) and $\Theta(x)$ is the Heaviside step function. The exact function $\gamma^\di(r)$ is given by \cite{BC87}
\beq
\gamma^\di_{\text{exact}}(r)=2^d\frac{{B}_{1-r^2/4}\left(\frac{d+1}{2},\frac{1}{2}\right)}{{B}\left(\frac{d+1}{2},\frac{1}{2}\right)}\Theta(2-r),
\label{12}
\eeq
$B_x(a,b)$ and $B(a,b)$ being the incomplete and complete beta functions, respectively \cite{AS72,OLBC10}.
In particular,
\beq
\label{13}
\gamma^{(1)}(r)=(2-r)\Theta(2-r),
\eeq
\beq
\label{14}
\gamma^{(3)}(r)=\frac{1}{2}\left(2-{r}\right)^2\left(4+{r}\right)\Theta(2-r).
\eeq
On the other hand, use of the approximation \eqref{6} yields
\beq
\label{15}
\gamma^\di(r)=\alpha^\di_0\lambda^\di_{1,0}\gamma^{(1)}(r)+[1-\alpha^\di_0]\lambda^\di_{3,0}\gamma^{(3)}(r),
\eeq
where
\begin{subequations}
\beq
\alpha^\di_0\equiv\lim_{\eeta\to 0}\alpha^\di(\eeta)=\frac{3}{34}\frac{50A^\di+22a^\di+25d-69}{2-a^\di},
\eeq
\beq
\lambda^\di_{1,0}\equiv\lim_{\eeta\to 0}\lambda^\di_1(\eeta)=2-a^\di,
\eeq
\beq
\lambda^\di_{3,0}\equiv\lim_{\eeta\to 0}\lambda^\di_3(\eeta)=\frac{2}{5}\left[2-a^\di\right].
\eeq
\end{subequations}
As seen from Fig.\ \ref{fig:gamma}, the values obtained from Eq.\ \eqref{15} are practically indistinguishable from the exact ones, Eq.\ \eqref{12}.

\begin{figure}
 \includegraphics[width=0.95\columnwidth]{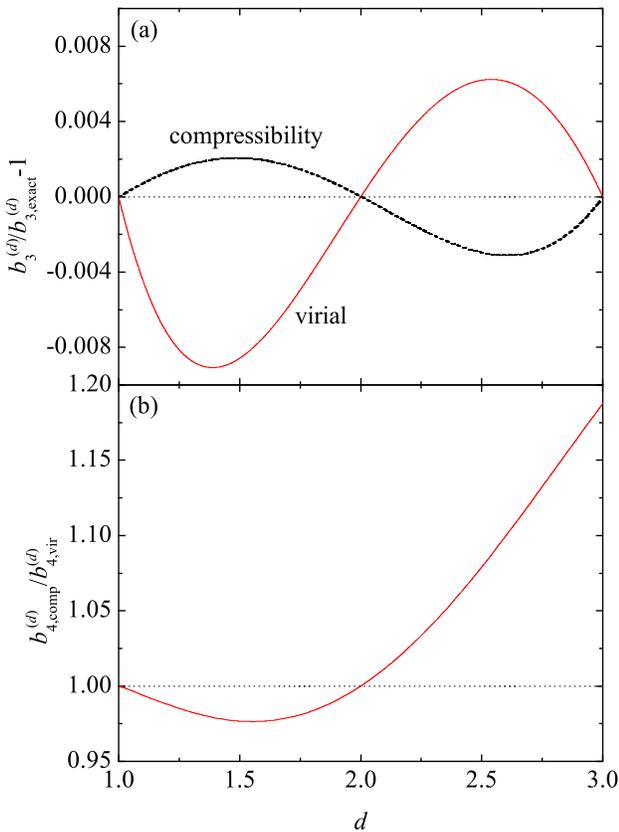}
\caption{(a) Relative differences  $b_{3,\text{vir}}^\di/b_{3,\text{exact}}^\di-1$ (solid line) and $b_{3,\text{comp}}^\di/b_{3,\text{exact}}^\di-1$ (dashed line) versus $d$. (b) Ratio $b_{4,\text{comp}}^\di/b_{4,\text{vir}}^\di$ versus $d$.}\label{fig:b3}
\end{figure}

Next, we consider the virial coefficients $b_n^\di$ defined by the power series
\beq
Z^\di(\eeta)=1+\sum_{n=2}^\infty b_n^\di \eeta^{n-1}.
\eeq
As can be seen from Eqs.\ \eqref{Z} and \eqref{11}, the second virial coefficient is $b_2^\di=2^{d-1}$. As for the third virial coefficient, its exact expression reads \cite{BC87}
\beq
b_{3,\text{exact}}^\di=2^{2d-1}\frac{B_{3/4}\left(\frac{d+1}{2},\frac{1}{2}\right)}{B\left(\frac{d+1}{2},\frac{1}{2}\right)}.
\label{b3ex}
\eeq
On the other hand, according to the approximation \eqref{15}, the  coefficients stemming from the virial route \eqref{Z} and from the compressibility route \eqref{chi_d} are
\begin{subequations}
\label{b3vir+comp}
\beq
b_{3,\text{vir}}^\di=2^{d-1}\left[2-a^\di\right],
\label{b3vir}
\eeq
\bal
b_{3,\text{comp}}^\di=&\frac{2^{d}}{3}\left\{2^d-\alpha_0^\di\lambda_{1,0}^\di\frac{2^{d+1}-d-2}{d+1}
-\left[1-\alpha_0^\di\right]\right.\nn
&\left.\times\lambda_{3,0}^\di\frac{3\times 2^{d+4}-5d^2-29d-48}{2(d+1)(d+3)}\right\}.
\label{b3comp}
\eal
\end{subequations}
The relative deviations of $b_{3,\text{vir}}^\di$ and $b_{3,\text{comp}}^\di$ from $b_{3,\text{exact}}^\di$ are plotted in Fig.\ \ref{fig:b3}(a). The  prescriptions \eqref{b3vir+comp}  are exact at $d=1$, $2$, and $3$. One can see that the highest deviations of $b_{3,\text{vir}}^\di$ from $b_{3,\text{exact}}^\di$ are $-0.9\%$ ($d=1.389$) and $+0.6\%$ ($d=2.539$). In the case of $b_{3,\text{comp}}^\di$ the highest deviations are $+0.2\%$ ($d=1.485$) and  $-0.3\%$ ($d=2.603$).

In respect to the fourth virial coefficient $b_4^\di$, although it is exactly known for integer values of $d$ \cite{CM04a,L05}, an analytic continuation to fractional $d$ is, to the best of our knowledge, not available. On the other hand, the fourth virial coefficients that follow from the virial route, $b_{4,\vir}^\di$, and from the compressibility route, $b_{4,\comp}^\di$, can easily be derived from Eqs.\ \eqref{Z} and \eqref{chi_d}, respectively. The ratio  $b_{4,\comp}^\di/b_{4,\vir}^\di$ is plotted in Fig.\ \ref{fig:b3}(b). We observe that, in contrast to what happens in the case of $b_3^\di$, one has $b_{4,\comp}^\di<b_{4,\vir}^\di$ for $1<d<2$ and $b_{4,\comp}^\di>b_{4,\vir}^\di$ for $2<d\leq 3$. Obviously, the PY value $b_{4,\comp}^{(3)}/b_{4,\vir}^{(3)}=19/16=1.1875 $ is recovered at $d=3$.

\begin{figure*}
 \includegraphics[width=1.8\columnwidth]{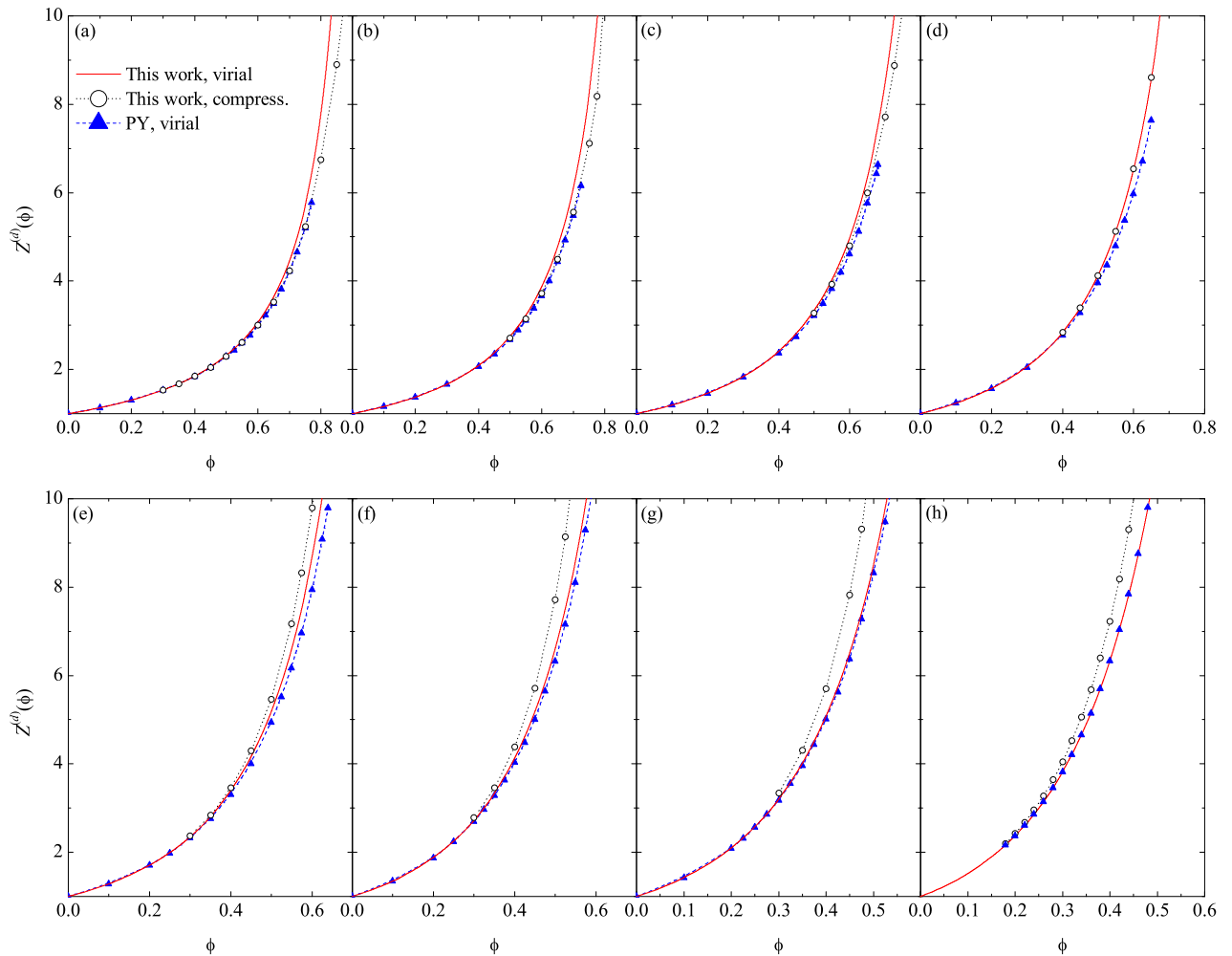}
\caption{Plot of the compressibility factor $Z^\di(\eeta)$ for  (a) $d=1.25$, (b) $d=1.5$, (c) $d=1.75$, (d) $d=2$, (e) $2.25$, (f) $d=2.5$, (g) $d=2.75$, and (h) $d=3$. The curves without symbols and the curves with circles represent the results obtained from our approach through the virial and compressibility routes, respectively. The curves  with triangles  correspond to $Z_\vir^\di(\eeta)$ as obtained from the numerical solution of the PY integral equation \cite{HSBL15}.}\label{fig:EOS}
\end{figure*}

\begin{figure}
 \includegraphics[width=.95\columnwidth]{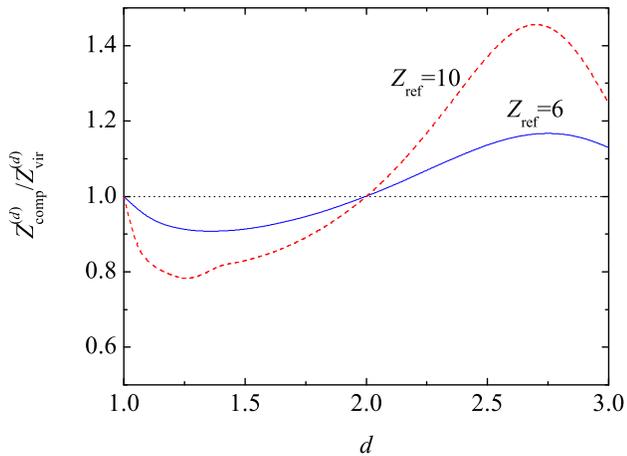}
\caption{Plot of the ratio $Z^\di_\comp/Z_\vir^\di$ as a function of $d$. For each value of $d$, a packing fraction $\eeta_{\text{ref}}$ is chosen such that $Z_\vir^\di(\eeta_{\text{ref}})=Z_{\text{ref}}$, where $Z_{\text{ref}}$ is a common value. The two curves correspond to $Z_{\text{ref}}=6$ and $Z_{\text{ref}}=10$.}\label{fig:Zref}
\end{figure}

\begin{figure}
 \includegraphics[width=0.95\columnwidth]{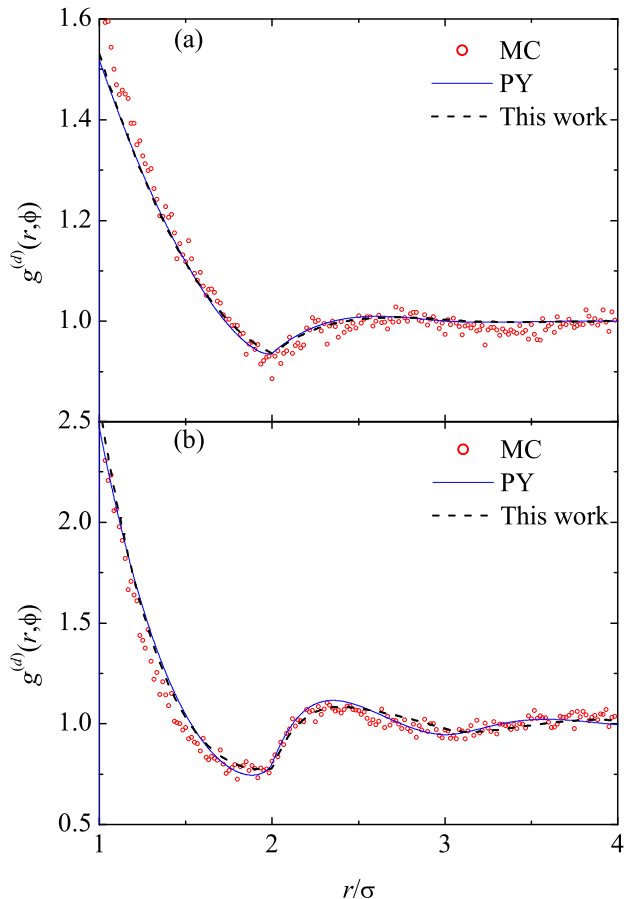}
\caption{Plot of $g^\di(r,\eeta)$ at (a) $\eeta=0.266$ and (b) $\eeta=0.487$  for $d=1.67659$. The symbols correspond to MC simulations \cite{HSBL15}, while the solid and dashed curves represent the numerical solution of the PY integral equation \cite{HSBL15} and the analytical approximation \eqref{6}, respectively.}\label{fig:RDF}
\end{figure}

\begin{figure}
 \includegraphics[width=0.95\columnwidth]{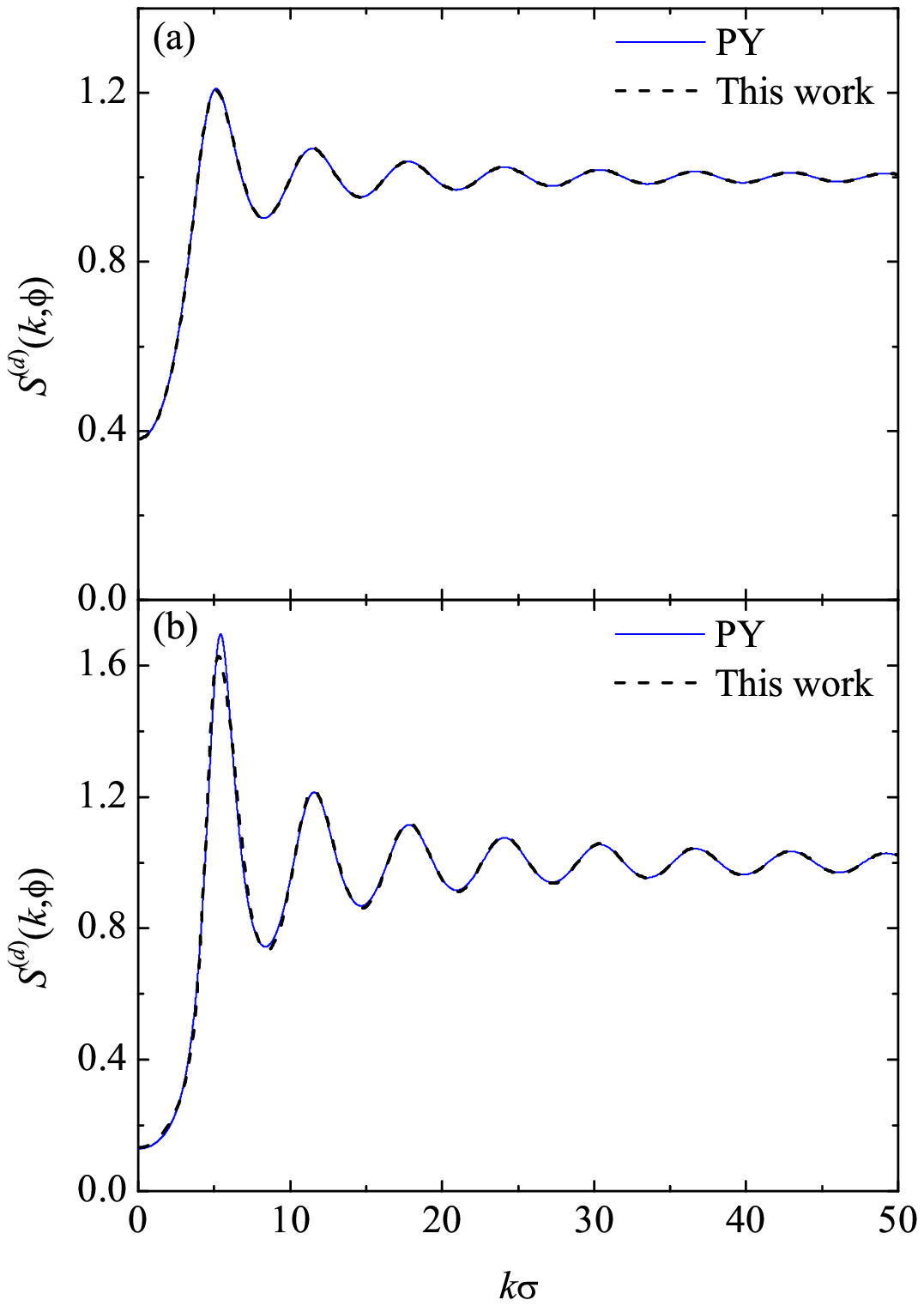}
\caption{Plot of $S^\di(k,\eeta)$ at (a) $\eeta=0.266$ and (b) $\eeta=0.487$  for $d=1.67659$. The solid and dashed curves represent the numerical solution of the PY integral equation \cite{note_16_04_2} and the approximation \eqref{S_kk}, respectively.}\label{fig:S_k}
\end{figure}

\section{Results at finite density}
\label{Results2}

Once we have studied the main properties of our proposal in the low-density regime, we turn our attention to the more important case of finite densities.

We first discuss the compressibility factor. Figure \ref{fig:EOS} compares $Z_\vir^\di(\eeta)$ and $Z_\comp^\di(\eeta)$, as predicted by our heuristic approximation, with $Z_\vir^\di(\eeta)$, as obtained from the numerical solution of the PY integral equation \cite{HSBL15,HAL14}.
It can be observed that the simple equation of state defined by the combination of Eqs.\ \eqref{g1-dD} and \eqref{Z} gives values close to (but slightly larger than) those obtained from the PY equation. Of course, both approximations coincide, by construction, at $d=3$. As for the compressibility-route function $Z_\comp^\di(\eeta)$ obtained from our approach, it practically coincides with $Z_\vir^\di(\eeta)$ up to $Z^\di\approx 4$. Thereafter, one has $Z_\comp^\di(\eeta)<Z_\vir^\di(\eeta)$ for $1<d<2$ and $Z_\comp^\di(\eeta)>Z_\vir^\di(\eeta)$ for $2<d\leq 3$, in qualitative agreement with Fig.\ \ref{fig:b3}(b).

To analyze the degree of thermodynamic inconsistency between $Z_\comp^\di(\eeta)$ and $Z_\vir^\di(\eeta)$ with more detail, we choose, for each $d$, a reference packing fraction $\eeta_{\text{ref}}$ such that $Z_\vir^\di(\eeta_{\text{ref}})=Z_{\text{ref}}$, where $Z_{\text{ref}}$ is a certain common value. Figure \ref{fig:Zref} shows the curves corresponding to $Z_{\text{ref}}=6$ and $Z_{\text{ref}}=10$. As can be observed, the inconsistency is typically smaller for $1<d<2$ than for $2<d<3$. In the latter interval, the highest discrepancy does not correspond to $d=3$ but to $d\approx 2.7$. Taking into account the known behavior at $d=3$, it is suggestive to speculate that the correct value of the compressibility factor lies approximately between $Z_\comp^\di(\eeta)$ and $Z_\vir^\di(\eeta)$, perhaps closer to the former than to the latter. This conjecture is supported by Fig.\ \ref{fig:b3}(a).

Let us  consider now the RDF. Figure \ref{fig:RDF} compares our simple approximation \eqref{6} against MC simulations and numerical solutions of the PY equation for the fractal dimensionality $d=1.67659$ and two packing fractions \cite{HSBL15}. We observe that at the lowest density ($\eeta=0.266$) the results obtained from the heuristic approach \eqref{6} are practically indistinguishable from the PY values, both describing fairly well the MC data. At a higher density ($\eeta=0.487$), however, small differences are visible, especially around the first minimum and the second maximum.
Although the noise in the MC data does not allow for a definite conclusion, a slightly better performance of Eq.\ \eqref{6} seems to be present.

The good agreement observed in Fig.\ \ref{fig:RDF} between our simple approach and the much more demanding numerical solution of the PY integral equation \cite{HAL14} is nicely confirmed by Fig.\ \ref{fig:S_k} at the level of the structure factor. Both theories yield practically indistinguishable results at $\eeta=0.266$. At $\eeta=0.487$, the only visible differences appear around the first peak and the first minimum (different from the one at $k=0$) of $S^\di(k,\eeta)$, where the values obtained from Eq.\ \eqref{S_kk} are slightly smaller than the PY ones.

\section{Concluding Remarks}
\label{concl}

Motivated by the recent work of Heinen \emph{et al.}\ \cite{HSBL15}, in this paper we have constructed a heuristic approximation for the RDF $g^\di(r,\eeta)$ and its associated structure factor $S^\di(k,\eeta)$ of hard-core fluids in fractal dimension ($1\leq d \leq 3$) [{cf.} Eqs.\ \eqref{6} and \eqref{S_kk}], which is exact for $d=1$ and reduces to the PY result for $d=3$. Also, using  $g^\di(r,\eeta)$, approximations for the compressibility factor $Z^\di(\eeta)$ of such fluids have been derived either analytically from the virial route [{cf.} Eq.\ \eqref{Z}, with the contact value $g_\con^\di(\eeta)$ given  by Eqs.\ \eqref{g1-dD}], or numerically from the compressibility route [{cf.} Eqs.\ \eqref{chi_d}]. A noteworthy aspect is that, by construction, the virial and compressibility routes are thermodynamically consistent in the cases $d=1$ (exact solution) and $d=2$ (Henderson's equation of state), while the two well-known PY equations of state are recovered in the case $d=3$. Moreover, we found that the approximate RDF to first order in density is practically indistinguishable from the exact one and the relative deviations of the third virial coefficients $b_{3,\text{vir}}^\di$ and $b_{3,\text{comp}}^\di$ (which are exact for $d=1$, $2$, and $3$) from $b_{3,\text{exact}}^\di$ are below $\pm1\%$. Further, the comparison between our results for $g^\di(r,\eeta)$ and the available MC and PY results \cite{HSBL15} for hard-core fluids in fractal dimension ($d=1.67659$), indicates that our proposal is much simpler and as accurate as the PY theory, and in some instances it shows a slightly better agreement with the MC data than the PY theory.

{As said} in Sec.\ \ref{sec2}, it would be possible to improve the accuracy of our proposal \eqref{6} by employing for the three-dimensional RDF $g^\thr(r,\eeta)$ a more consistent and reliable form \cite{S16,YS91,YHS96,HYS08} than the PY one. While this alternative implementation is not expected to have a relevant impact for fractal dimensions $1<d<2$, it might be interesting to explore it in the range $2<d<3$, provided computer simulations and/or numerical solutions of integral equations were available. Another possible extension of our work {in the same range of values for the fractal dimension} is the case of mixtures, where again the exact solution for $d=1$ and the PY solution for $d=3$ are known \cite{HC04,L64,S16}.

{Although our proposal has been devised explicitly for $1< d< 3$, in principle one could attempt to extrapolate it for fractal dimensions slightly higher than $d=3$. Of course it should be expected that the quality of the agreement will deteriorate as $d$ increases, especially beyond the low-density regime. As an illustration that this expectation is fulfilled, we find that, for $d=4$, our approximation yields $b_{3,\text{vir}}^{(4)}=30.4638$, $b_{3,\text{comp}}^{(4)}=34.3734$, $b_{4,\text{vir}}^{(4)}=52.9276$, and $b_{4,\text{comp}}^{(4)}=64.3768$, while the exact results are $b_{3,\text{exact}}^{(4)}=32.4058$ and $b_{4,\text{exact}}^{(4)}=77.7452$. In fact, for $3< d< 5$, a much better approximation should be obtained by following a similar interpolation procedure but using the known PY results for $g^{(3)}(r,\eeta)$ and $g^{(5)}(r,\eeta)$ \cite{FI81,L84,RS07} instead of $g^{(1)}(r,\eeta)$ and $g^{(3)}(r,\eeta)$, as done here. Finally, as far as possible generalizations to particle interactions other than hard-core within the range $1< d< 3$ are concerned, the case that immediately comes to mind is the one corresponding to sticky hard spheres. In such a case the exact RDF is known for $d=1$ \cite{YS93a,LNP_book_note_13_08,S16} and the PY result for $d=3$ is also available \cite{B68,B74,S16}.}

\begin{acknowledgments}
We want to thank M. Heinen  for supplying simulation MC data and numerical results of the PY theory, as well as for a fruitful exchange of correspondence. We also acknowledge the financial support of the Ministerio de  Econom\'ia y Competitividad (Spain) through Grant No.\ FIS2013-42840-P and  the Junta de Extremadura (Spain) through Grant No.\ GR15104 (partially financed by FEDER funds).
\end{acknowledgments}

\appendix*
\section{Radial distribution function of hard rods and hard spheres}
\subsection{Hard rods}
In the one-dimensional case ($d=1$), the Laplace transform
\beq
G^\on(s,\eeta)=\int_0^\infty \dd r\, e^{-rs} g^\on(r,\eeta)
\label{A1}
\eeq
is exactly given by \cite{HC04,S16}
\beq
G^\on(s,\eeta)=\frac{1}{\eeta}\frac{e^{-s}}{1+s(1-\eeta)/\eeta-e^{-s}},
\label{A2}
\eeq
where again we have set $\sigma=1$ as length unit.
Expanding $G^\on(s,\eeta)$ in powers of $e^{-s}$, it is easy to perform an inverse Laplace transform term by term to obtain
\bal
g^\on(r,\eeta)=&\frac{1}{\eeta}\sum_{\ell=1}^\infty \left(\frac{\eeta}{1-\eeta}\right)^\ell\frac{(r-\ell)^{\ell-1}}{(\ell-1)!}\nn
&\times e^{-(r-\ell)\eeta/(1-\eeta)}\Theta(r-\ell).
\label{A3}
\eal
This representation can be truncated at $\ell=\ell_c$ if only the range $1\leq r\leq \ell_c+1$ is relevant. On the other hand, the tail of the total correlation function is in principle necessary to complement \eqref{A3} in the evaluation of the integrals in \eqref{chi_dd} and \eqref{I_k}. The asymptotic tail of $h^\on(r,\eeta)$ is \cite{PT72}
\beq
h^\on(r,\eeta)\approx K^\on(\eeta)e^{-\kappa^\on(\eeta)r}\cos[\omega^\on(\eeta)r+\delta^\on(\eeta)],
\label{A4}
\eeq
where $s_{\pm}=-\kappa^\on\pm i\omega^\on$ are the two complex conjugate poles of $G^\on(s)$ with a real part closest to the origin and $\frac{1}{2}K^\on e^{i\delta^\on}$ is the associated residue.
More explicitly, $\kappa^\on$ and $\omega^\on$ are roots of the coupled set of transcendental equations
\begin{subequations}
\bal
  1-\frac{1-\eeta}{\eeta}\kappa^\on=&e^{\kappa^\on}\cos\omega^\on,
\label{A5a}
\\
-\frac{1-\eeta}{\eeta}\omega^\on=&e^{\kappa^\on}\sin\omega^\on.
\label{A5b}
\eal
\end{subequations}
Once $s_{\pm}=-\kappa^\on\pm i\omega^\on$ are known, the amplitude and phase are
\begin{subequations}
\bal
K^\on=&2\left|1+\frac{1-\eeta}{\eeta+(1-\eeta)s_\pm}\right|^{-1},
\label{A6a}
\\
\delta^\on=&-\arg\left[1+\frac{1-\eeta}{\eeta+(1-\eeta)s_\pm}\right].
\label{A6b}
\eal
\end{subequations}

\subsection{Hard spheres}
The PY solution for the three-dimensional case has a structure reminiscent of that of the one-dimensional case. First, the Laplace transform
\beq
G^\thr(s,\eeta)=\int_0^\infty \dd r\, e^{-rs} rg^\thr(r,\eeta)
\label{A7}
\eeq
is introduced. The PY solution is then found to be \cite{W63,W64,S16}
\beq
G^\thr(s,\eeta)={s}\frac{F(s,\eeta)e^{-s}}{1+12\eeta F(s,\eeta)e^{-s}},
\label{A8}
\eeq
where
\beq
F(s,\eeta)=-\frac{1}{12\eeta}\frac{1+L_1(\eeta) s}{1+S_1(\eeta) s+S_2(\eeta) s^2+S_3(\eeta) s^3}
\eeq
with the coefficients
\begin{subequations}
\bal
L_1(\eeta)=&\frac{1+\eeta/2}{1+2\eeta},\quad S_1(\eeta)=-\frac{3}{2}\frac{\eeta}{1+2\eeta},
\\
S_2(\eeta)=&-\frac{1}{2}\frac{1-\eeta}{1+2\eeta},\quad
     S_3(\eeta)=-\frac{1}{12\eeta}\frac{(1-\eeta)^2}{1+2\eeta}.
\eal
\end{subequations}
Again, a formal expansion of Eq.\ \eqref{A8} in powers of $e^{-s}$ allows one to write
\beq
g^\thr(r,\eeta)=\frac{1}{r}\sum_{\ell=1}^\infty \left(-12\eeta\right)^{\ell-1}{\Psi}_{\ell}(r-\ell,\eeta){\Theta(r-\ell)},
\label{A9}
    \eeq
where
    \beq
{\Psi}_\ell(r,\eeta)=\sum_{j=1}^\ell \frac{\sum_{i=1}^3 a_{\ell j}^{(i)}(\eeta)e^{s_i(\eeta) r}}{(\ell-j)!(j-1)!}r^{\ell-j},
\eeq
\beq
a_{\ell j}^{(i)}(\eeta)=\lim_{s\to s_i(\eeta)}\left(\frac{\partial}{\partial s}\right)^{j-1}\left\{s\left[(s-s_i(\eeta))F(s,\eeta)\right]^\ell\right\}.
\eeq
Here, $s_i(\eeta)$, ($i=1,2,3$) are the three roots of the cubic equation $1+S_1(\eeta) s+S_2(\eeta) s^2+S_3(\eeta) s^3=0$. As in the case of Eq.\ \eqref{A3}, the summation in Eq.\ \eqref{A9} can be truncated at $\ell=\ell_c$ to obtain the RDF within the interval $1\leq r\leq \ell_c+1$. This can be complemented by the asymptotic tail of the total correlation function
\cite{PT72,TS77},
\beq
h^\thr(r,\eeta)\approx \frac{K^\thr(\eeta)}{r}e^{-\kappa^\thr(\eeta)r}\cos[\omega^\thr(\eeta)r+\delta^\thr(\eeta)],
\label{A10}
\eeq
where $s_{\pm}=-\kappa^\thr\pm i\omega^\thr$ are the two complex conjugate poles of $G^\thr(s)$ with a real part closest to the origin, $\frac{1}{2}K^\thr e^{i\delta^\thr}$ being the associated residue. Thus, the transcendental equation for $s_\pm$ is
\beq
\left(1+L_1s_\pm\right)e^{-s_\pm}=1+S_1 s_\pm+S_2 s_\pm^2+S_3 s_\pm^3,
\eeq
while the residue is
\beq
\text{Res}=\frac{s_\pm\left(1+L_1 s_\pm\right)/12\eeta}{L_1(s_\pm-1)-1-e^{s_\pm}\left(S_1+2S_2 s_\pm+3S_3 s_\pm^2\right)},
\eeq
so that
 \beq
 K^\thr= 2|\text{Res}|,\quad
 \delta^\thr=\arg(\text{Res}).
 \eeq

\bibliography{D:/Dropbox/Public/bib_files/liquid}

\begin{thebibliography}{41}%
\makeatletter
\providecommand \@ifxundefined [1]{%
 \@ifx{#1\undefined}
}%
\providecommand \@ifnum [1]{%
 \ifnum #1\expandafter \@firstoftwo
 \else \expandafter \@secondoftwo
 \fi
}%
\providecommand \@ifx [1]{%
 \ifx #1\expandafter \@firstoftwo
 \else \expandafter \@secondoftwo
 \fi
}%
\providecommand \natexlab [1]{#1}%
\providecommand \enquote  [1]{``#1''}%
\providecommand \bibnamefont  [1]{#1}%
\providecommand \bibfnamefont [1]{#1}%
\providecommand \citenamefont [1]{#1}%
\providecommand \href@noop [0]{\@secondoftwo}%
\providecommand \href [0]{\begingroup \@sanitize@url \@href}%
\providecommand \@href[1]{\@@startlink{#1}\@@href}%
\providecommand \@@href[1]{\endgroup#1\@@endlink}%
\providecommand \@sanitize@url [0]{\catcode `\\12\catcode `\$12\catcode
  `\&12\catcode `\#12\catcode `\^12\catcode `\_12\catcode `\%12\relax}%
\providecommand \@@startlink[1]{}%
\providecommand \@@endlink[0]{}%
\providecommand \url  [0]{\begingroup\@sanitize@url \@url }%
\providecommand \@url [1]{\endgroup\@href {#1}{\urlprefix }}%
\providecommand \urlprefix  [0]{URL }%
\providecommand \Eprint [0]{\href }%
\providecommand \doibase [0]{http://dx.doi.org/}%
\providecommand \selectlanguage [0]{\@gobble}%
\providecommand \bibinfo  [0]{\@secondoftwo}%
\providecommand \bibfield  [0]{\@secondoftwo}%
\providecommand \translation [1]{[#1]}%
\providecommand \BibitemOpen [0]{}%
\providecommand \bibitemStop [0]{}%
\providecommand \bibitemNoStop [0]{.\EOS\space}%
\providecommand \EOS [0]{\spacefactor3000\relax}%
\providecommand \BibitemShut  [1]{\csname bibitem#1\endcsname}%
\let\auto@bib@innerbib\@empty
\bibitem [{\citenamefont {Charbonneau}\ \emph {et~al.}(2011)\citenamefont
  {Charbonneau}, \citenamefont {Ikeda}, \citenamefont {Parisi},\ and\
  \citenamefont {Zamponi}}]{CIPZ2011}%
  \BibitemOpen
  \bibfield  {author} {\bibinfo {author} {\bibfnamefont {P.}~\bibnamefont
  {Charbonneau}}, \bibinfo {author} {\bibfnamefont {A.}~\bibnamefont {Ikeda}},
  \bibinfo {author} {\bibfnamefont {G.}~\bibnamefont {Parisi}}, \ and\ \bibinfo
  {author} {\bibfnamefont {F.}~\bibnamefont {Zamponi}},\ }\href@noop {}
  {\bibfield  {journal} {\bibinfo  {journal} {Phys. Rev. Lett.}\ }\textbf
  {\bibinfo {volume} {107}},\ \bibinfo {pages} {185702} (\bibinfo {year}
  {2011})}\BibitemShut {NoStop}%
\bibitem [{\citenamefont {L\"{o}wen}(2009)}]{L09}%
  \BibitemOpen
  \bibfield  {author} {\bibinfo {author} {\bibfnamefont {H.}~\bibnamefont
  {L\"{o}wen}},\ }\href@noop {} {\bibfield  {journal} {\bibinfo  {journal} {J.
  Phys.: Condens. Matter}\ }\textbf {\bibinfo {volume} {21}},\ \bibinfo {pages}
  {474203} (\bibinfo {year} {2009})}\BibitemShut {NoStop}%
\bibitem [{\citenamefont {Franosch}\ \emph {et~al.}(2012)\citenamefont
  {Franosch}, \citenamefont {Lang},\ and\ \citenamefont {Schilling}}]{FLS12}%
  \BibitemOpen
  \bibfield  {author} {\bibinfo {author} {\bibfnamefont {T.}~\bibnamefont
  {Franosch}}, \bibinfo {author} {\bibfnamefont {S.}~\bibnamefont {Lang}}, \
  and\ \bibinfo {author} {\bibfnamefont {R.}~\bibnamefont {Schilling}},\
  }\href@noop {} {\bibfield  {journal} {\bibinfo  {journal} {Phys. Rev. Lett.}\
  }\textbf {\bibinfo {volume} {109}},\ \bibinfo {pages} {240601} (\bibinfo
  {year} {2012})}\BibitemShut {NoStop}%
\bibitem [{\citenamefont {Franosch}\ \emph {et~al.}(2013)\citenamefont
  {Franosch}, \citenamefont {Lang},\ and\ \citenamefont {Schilling}}]{FLS13}%
  \BibitemOpen
  \bibfield  {author} {\bibinfo {author} {\bibfnamefont {T.}~\bibnamefont
  {Franosch}}, \bibinfo {author} {\bibfnamefont {S.}~\bibnamefont {Lang}}, \
  and\ \bibinfo {author} {\bibfnamefont {R.}~\bibnamefont {Schilling}},\
  }\href@noop {} {\bibfield  {journal} {\bibinfo  {journal} {Phys. Rev. Lett.}\
  }\textbf {\bibinfo {volume} {110}},\ \bibinfo {pages} {059901(E)} (\bibinfo
  {year} {2013})}\BibitemShut {NoStop}%
\bibitem [{\citenamefont {Heinen}\ \emph {et~al.}(2015)\citenamefont {Heinen},
  \citenamefont {Schnyder}, \citenamefont {Brady},\ and\ \citenamefont
  {L\"owen}}]{HSBL15}%
  \BibitemOpen
  \bibfield  {author} {\bibinfo {author} {\bibfnamefont {M.}~\bibnamefont
  {Heinen}}, \bibinfo {author} {\bibfnamefont {S.~K.}\ \bibnamefont
  {Schnyder}}, \bibinfo {author} {\bibfnamefont {J.~F.}\ \bibnamefont {Brady}},
  \ and\ \bibinfo {author} {\bibfnamefont {H.}~\bibnamefont {L\"owen}},\
  }\href@noop {} {\bibfield  {journal} {\bibinfo  {journal} {Phys. Rev. Lett.}\
  }\textbf {\bibinfo {volume} {115}},\ \bibinfo {pages} {097801} (\bibinfo
  {year} {2015})}\BibitemShut {NoStop}%
\bibitem [{\citenamefont {Yuste}\ and\ \citenamefont
  {Santos}(1993{\natexlab{a}})}]{YS93c}%
  \BibitemOpen
  \bibfield  {author} {\bibinfo {author} {\bibfnamefont {S.~B.}\ \bibnamefont
  {Yuste}}\ and\ \bibinfo {author} {\bibfnamefont {A.}~\bibnamefont {Santos}},\
  }\href@noop {} {\bibfield  {journal} {\bibinfo  {journal} {J. Chem. Phys.}\
  }\textbf {\bibinfo {volume} {99}},\ \bibinfo {pages} {2020} (\bibinfo {year}
  {1993}{\natexlab{a}})}\BibitemShut {NoStop}%
\bibitem [{\citenamefont {Herzfeld}\ and\ \citenamefont
  {Goeppert-Mayer}(1934)}]{HGM34}%
  \BibitemOpen
  \bibfield  {author} {\bibinfo {author} {\bibfnamefont {K.~F.}\ \bibnamefont
  {Herzfeld}}\ and\ \bibinfo {author} {\bibfnamefont {M.}~\bibnamefont
  {Goeppert-Mayer}},\ }\href@noop {} {\bibfield  {journal} {\bibinfo  {journal}
  {J. Chem. Phys.}\ }\textbf {\bibinfo {volume} {2}},\ \bibinfo {pages} {38}
  (\bibinfo {year} {1934})}\BibitemShut {NoStop}%
\bibitem [{\citenamefont {Tonks}(1936)}]{T36}%
  \BibitemOpen
  \bibfield  {author} {\bibinfo {author} {\bibfnamefont {L.}~\bibnamefont
  {Tonks}},\ }\href@noop {} {\bibfield  {journal} {\bibinfo  {journal} {Phys.
  Rev.}\ }\textbf {\bibinfo {volume} {50}},\ \bibinfo {pages} {955} (\bibinfo
  {year} {1936})}\BibitemShut {NoStop}%
\bibitem [{\citenamefont {Heying}\ and\ \citenamefont {Corti}(2004)}]{HC04}%
  \BibitemOpen
  \bibfield  {author} {\bibinfo {author} {\bibfnamefont {M.}~\bibnamefont
  {Heying}}\ and\ \bibinfo {author} {\bibfnamefont {D.~S.}\ \bibnamefont
  {Corti}},\ }\href@noop {} {\bibfield  {journal} {\bibinfo  {journal} {Fluid
  Phase Equil.}\ }\textbf {\bibinfo {volume} {220}},\ \bibinfo {pages} {85}
  (\bibinfo {year} {2004})}\BibitemShut {NoStop}%
\bibitem [{\citenamefont {Santos}(2014)}]{S14}%
  \BibitemOpen
  \bibfield  {author} {\bibinfo {author} {\bibfnamefont {A.}~\bibnamefont
  {Santos}},\ }in\ \href@noop {} {\emph {\bibinfo {booktitle} {5th Warsaw
  School of Statistical Physics}}},\ \bibinfo {editor} {edited by\ \bibinfo
  {editor} {\bibfnamefont {B.}~\bibnamefont {Cichocki}}, \bibinfo {editor}
  {\bibfnamefont {M.}~\bibnamefont {Napi\'orkowski}}, \ and\ \bibinfo {editor}
  {\bibfnamefont {J.}~\bibnamefont {Piasecki}}}\ (\bibinfo  {publisher} {Warsaw
  University Press},\ \bibinfo {address} {Warsaw},\ \bibinfo {year} {2014})\
  \bibinfo {note}
  {\href{arXiv:1310.5578}{http://arxiv.org/abs/1310.5578}}\BibitemShut
  {NoStop}%
\bibitem [{\citenamefont {Santos}(2016)}]{S16}%
  \BibitemOpen
  \bibfield  {author} {\bibinfo {author} {\bibfnamefont {A.}~\bibnamefont
  {Santos}},\ }\href@noop {} {\emph {\bibinfo {title} {{A Concise Course on the
  Theory of Classical Liquids. Basics and Selected Topics}}}},\ \bibinfo
  {series} {Lecture Notes in Physics}, Vol.\ \bibinfo {volume} {923}\ (\bibinfo
   {publisher} {Springer},\ \bibinfo {year} {2016})\BibitemShut {NoStop}%
\bibitem [{\citenamefont {Santos}(2015)}]{LNP_book_note_15_06_1}%
  \BibitemOpen
  \bibfield  {author} {\bibinfo {author} {\bibfnamefont {A.}~\bibnamefont
  {Santos}},\ }\href@noop {} {} (\bibinfo {year} {2015}),\ \bibinfo {note}
  {``Radial Distribution Functions for Nonadditive Hard-Rod Mixtures'', Wolfram
  Demonstrations Project,
  \url{http://demonstrations.wolfram.com/RadialDistributionFunctionsForNonadditiveHardRodMixtures/}}\BibitemShut
  {NoStop}%
\bibitem [{\citenamefont {Thiele}(1963)}]{T63}%
  \BibitemOpen
  \bibfield  {author} {\bibinfo {author} {\bibfnamefont {E.}~\bibnamefont
  {Thiele}},\ }\href@noop {} {\bibfield  {journal} {\bibinfo  {journal} {J.
  Chem. Phys.}\ }\textbf {\bibinfo {volume} {39}},\ \bibinfo {pages} {474}
  (\bibinfo {year} {1963})}\BibitemShut {NoStop}%
\bibitem [{\citenamefont {Wertheim}(1963)}]{W63}%
  \BibitemOpen
  \bibfield  {author} {\bibinfo {author} {\bibfnamefont {M.~S.}\ \bibnamefont
  {Wertheim}},\ }\href@noop {} {\bibfield  {journal} {\bibinfo  {journal}
  {Phys. Rev. Lett.}\ }\textbf {\bibinfo {volume} {10}},\ \bibinfo {pages}
  {321} (\bibinfo {year} {1963})}\BibitemShut {NoStop}%
\bibitem [{\citenamefont {Wertheim}(1964)}]{W64}%
  \BibitemOpen
  \bibfield  {author} {\bibinfo {author} {\bibfnamefont {M.~S.}\ \bibnamefont
  {Wertheim}},\ }\href@noop {} {\bibfield  {journal} {\bibinfo  {journal} {J.
  Math. Phys.}\ }\textbf {\bibinfo {volume} {5}},\ \bibinfo {pages} {643}
  (\bibinfo {year} {1964})}\BibitemShut {NoStop}%
\bibitem [{\citenamefont {Hansen}\ and\ \citenamefont {McDonald}(2006)}]{HM06}%
  \BibitemOpen
  \bibfield  {author} {\bibinfo {author} {\bibfnamefont {J.-P.}\ \bibnamefont
  {Hansen}}\ and\ \bibinfo {author} {\bibfnamefont {I.~R.}\ \bibnamefont
  {McDonald}},\ }\href@noop {} {\emph {\bibinfo {title} {{Theory of Simple
  Liquids}}}},\ \bibinfo {edition} {3rd}\ ed.\ (\bibinfo  {publisher}
  {Academic},\ \bibinfo {address} {London},\ \bibinfo {year}
  {2006})\BibitemShut {NoStop}%
\bibitem [{\citenamefont {Santos}(2013)}]{LNP_book_note_13_10}%
  \BibitemOpen
  \bibfield  {author} {\bibinfo {author} {\bibfnamefont {A.}~\bibnamefont
  {Santos}},\ }\href@noop {} {} (\bibinfo {year} {2013}),\ \bibinfo {note}
  {``Radial Distribution Function for Hard Spheres'', Wolfram Demonstrations
  Project,
  \url{http://demonstrations.wolfram.com/RadialDistributionFunctionForHardSpheres/}}\BibitemShut
  {NoStop}%
\bibitem [{\citenamefont {Reiss}\ \emph {et~al.}(1959)\citenamefont {Reiss},
  \citenamefont {Frisch},\ and\ \citenamefont {Lebowitz}}]{RFL59}%
  \BibitemOpen
  \bibfield  {author} {\bibinfo {author} {\bibfnamefont {H.}~\bibnamefont
  {Reiss}}, \bibinfo {author} {\bibfnamefont {H.~L.}\ \bibnamefont {Frisch}}, \
  and\ \bibinfo {author} {\bibfnamefont {J.~L.}\ \bibnamefont {Lebowitz}},\
  }\href@noop {} {\bibfield  {journal} {\bibinfo  {journal} {J. Chem. Phys.}\
  }\textbf {\bibinfo {volume} {31}},\ \bibinfo {pages} {369} (\bibinfo {year}
  {1959})}\BibitemShut {NoStop}%
\bibitem [{\citenamefont {Helfand}\ \emph {et~al.}(1961)\citenamefont
  {Helfand}, \citenamefont {Frisch},\ and\ \citenamefont {Lebowitz}}]{HFL61}%
  \BibitemOpen
  \bibfield  {author} {\bibinfo {author} {\bibfnamefont {E.}~\bibnamefont
  {Helfand}}, \bibinfo {author} {\bibfnamefont {H.~L.}\ \bibnamefont {Frisch}},
  \ and\ \bibinfo {author} {\bibfnamefont {J.~L.}\ \bibnamefont {Lebowitz}},\
  }\href@noop {} {\bibfield  {journal} {\bibinfo  {journal} {J. Chem. Phys.}\
  }\textbf {\bibinfo {volume} {34}},\ \bibinfo {pages} {1037} (\bibinfo {year}
  {1961})}\BibitemShut {NoStop}%
\bibitem [{\citenamefont {Henderson}(1975)}]{H75}%
  \BibitemOpen
  \bibfield  {author} {\bibinfo {author} {\bibfnamefont {D.}~\bibnamefont
  {Henderson}},\ }\href@noop {} {\bibfield  {journal} {\bibinfo  {journal}
  {Mol. Phys.}\ }\textbf {\bibinfo {volume} {30}},\ \bibinfo {pages} {971}
  (\bibinfo {year} {1975})}\BibitemShut {NoStop}%
\bibitem [{\citenamefont {Yuste}\ and\ \citenamefont {Santos}(1991)}]{YS91}%
  \BibitemOpen
  \bibfield  {author} {\bibinfo {author} {\bibfnamefont {S.~B.}\ \bibnamefont
  {Yuste}}\ and\ \bibinfo {author} {\bibfnamefont {A.}~\bibnamefont {Santos}},\
  }\href@noop {} {\bibfield  {journal} {\bibinfo  {journal} {Phys. Rev. A}\
  }\textbf {\bibinfo {volume} {43}},\ \bibinfo {pages} {5418} (\bibinfo {year}
  {1991})}\BibitemShut {NoStop}%
\bibitem [{\citenamefont {Yuste}\ \emph {et~al.}(1996)\citenamefont {Yuste},
  \citenamefont {{L\'opez de Haro}},\ and\ \citenamefont {Santos}}]{YHS96}%
  \BibitemOpen
  \bibfield  {author} {\bibinfo {author} {\bibfnamefont {S.~B.}\ \bibnamefont
  {Yuste}}, \bibinfo {author} {\bibfnamefont {M.}~\bibnamefont {{L\'opez de
  Haro}}}, \ and\ \bibinfo {author} {\bibfnamefont {A.}~\bibnamefont
  {Santos}},\ }\href@noop {} {\bibfield  {journal} {\bibinfo  {journal} {Phys.
  Rev. E}\ }\textbf {\bibinfo {volume} {53}},\ \bibinfo {pages} {4820}
  (\bibinfo {year} {1996})}\BibitemShut {NoStop}%
\bibitem [{\citenamefont {{L\'opez de Haro}}\ \emph {et~al.}(2008)\citenamefont
  {{L\'opez de Haro}}, \citenamefont {Yuste},\ and\ \citenamefont
  {Santos}}]{HYS08}%
  \BibitemOpen
  \bibfield  {author} {\bibinfo {author} {\bibfnamefont {M.}~\bibnamefont
  {{L\'opez de Haro}}}, \bibinfo {author} {\bibfnamefont {S.~B.}\ \bibnamefont
  {Yuste}}, \ and\ \bibinfo {author} {\bibfnamefont {A.}~\bibnamefont
  {Santos}},\ }in\ \href@noop {} {\emph {\bibinfo {booktitle} {{Theory and
  Simulation of Hard-Sphere Fluids and Related Systems}}}},\ \bibinfo {series}
  {Lecture Notes in Physics}, Vol.\ \bibinfo {volume} {753},\ \bibinfo {editor}
  {edited by\ \bibinfo {editor} {\bibfnamefont {A.}~\bibnamefont {Mulero}}}\
  (\bibinfo  {publisher} {Springer-Verlag},\ \bibinfo {address} {Berlin},\
  \bibinfo {year} {2008})\ pp.\ \bibinfo {pages} {183--245}\BibitemShut
  {NoStop}%
\bibitem [{not(2016{\natexlab{a}})}]{note_16_04_1}%
  \BibitemOpen
  \href@noop {} {} (\bibinfo {year} {2016}{\natexlab{a}}),\ \bibinfo {note}
  {{S}ee Supplemental Material at \url{http://link.aps.org/supplemental/
  10.1103/PhysRevE.93.062126} for a Mathematica code with all the expressions
  of this work}\BibitemShut {NoStop}%
\bibitem [{\citenamefont {Baus}\ and\ \citenamefont {Colot}(1987)}]{BC87}%
  \BibitemOpen
  \bibfield  {author} {\bibinfo {author} {\bibfnamefont {M.}~\bibnamefont
  {Baus}}\ and\ \bibinfo {author} {\bibfnamefont {J.~L.}\ \bibnamefont
  {Colot}},\ }\href@noop {} {\bibfield  {journal} {\bibinfo  {journal} {Phys.
  Rev. A}\ }\textbf {\bibinfo {volume} {36}},\ \bibinfo {pages} {3912}
  (\bibinfo {year} {1987})}\BibitemShut {NoStop}%
\bibitem [{\citenamefont {Abramowitz}\ and\ \citenamefont
  {Stegun}(1972)}]{AS72}%
  \BibitemOpen
  \bibinfo {editor} {\bibfnamefont {M.}~\bibnamefont {Abramowitz}}\ and\
  \bibinfo {editor} {\bibfnamefont {I.~A.}\ \bibnamefont {Stegun}},\ eds.,\
  \href@noop {} {\emph {\bibinfo {title} {{Handbook of Mathematical
  Functions}}}}\ (\bibinfo  {publisher} {Dover},\ \bibinfo {address} {New
  York},\ \bibinfo {year} {1972})\BibitemShut {NoStop}%
\bibitem [{\citenamefont {Olver}\ \emph {et~al.}(2010)\citenamefont {Olver},
  \citenamefont {Lozier}, \citenamefont {Boisvert},\ and\ \citenamefont
  {Clark}}]{OLBC10}%
  \BibitemOpen
  \bibinfo {editor} {\bibfnamefont {F.~W.~J.}\ \bibnamefont {Olver}}, \bibinfo
  {editor} {\bibfnamefont {D.~W.}\ \bibnamefont {Lozier}}, \bibinfo {editor}
  {\bibfnamefont {R.~F.}\ \bibnamefont {Boisvert}}, \ and\ \bibinfo {editor}
  {\bibfnamefont {C.~W.}\ \bibnamefont {Clark}},\ eds.,\ \href@noop {} {\emph
  {\bibinfo {title} {NIST Handbook of Mathematical Functions}}}\ (\bibinfo
  {publisher} {Cambridge University Press},\ \bibinfo {address} {New York},\
  \bibinfo {year} {2010})\BibitemShut {NoStop}%
\bibitem [{\citenamefont {Clisby}\ and\ \citenamefont {McCoy}(2004)}]{CM04a}%
  \BibitemOpen
  \bibfield  {author} {\bibinfo {author} {\bibfnamefont {N.}~\bibnamefont
  {Clisby}}\ and\ \bibinfo {author} {\bibfnamefont {B.~M.}\ \bibnamefont
  {McCoy}},\ }\href@noop {} {\bibfield  {journal} {\bibinfo  {journal} {J.
  Stat. Phys.}\ }\textbf {\bibinfo {volume} {114}},\ \bibinfo {pages} {1343}
  (\bibinfo {year} {2004})}\BibitemShut {NoStop}%
\bibitem [{\citenamefont {Lyberg}(2005)}]{L05}%
  \BibitemOpen
  \bibfield  {author} {\bibinfo {author} {\bibfnamefont {I.}~\bibnamefont
  {Lyberg}},\ }\href@noop {} {\bibfield  {journal} {\bibinfo  {journal} {J.
  Stat. Phys.}\ }\textbf {\bibinfo {volume} {119}},\ \bibinfo {pages} {747}
  (\bibinfo {year} {2005})}\BibitemShut {NoStop}%
\bibitem [{not(2016{\natexlab{b}})}]{note_16_04_2}%
  \BibitemOpen
  \href@noop {} {} (\bibinfo {year} {2016}{\natexlab{b}}),\ \bibinfo {note}
  {{M}. Heinen, private communication}\BibitemShut {NoStop}%
\bibitem [{\citenamefont {Heinen}\ \emph {et~al.}(2014)\citenamefont {Heinen},
  \citenamefont {Allahyarov},\ and\ \citenamefont {L\"owen}}]{HAL14}%
  \BibitemOpen
  \bibfield  {author} {\bibinfo {author} {\bibfnamefont {M.}~\bibnamefont
  {Heinen}}, \bibinfo {author} {\bibfnamefont {E.}~\bibnamefont {Allahyarov}},
  \ and\ \bibinfo {author} {\bibfnamefont {H.}~\bibnamefont {L\"owen}},\
  }\href@noop {} {\bibfield  {journal} {\bibinfo  {journal} {J. Comput. Chem.}\
  }\textbf {\bibinfo {volume} {35}},\ \bibinfo {pages} {275} (\bibinfo {year}
  {2014})}\BibitemShut {NoStop}%
\bibitem [{\citenamefont {Lebowitz}(1964)}]{L64}%
  \BibitemOpen
  \bibfield  {author} {\bibinfo {author} {\bibfnamefont {J.~L.}\ \bibnamefont
  {Lebowitz}},\ }\href@noop {} {\bibfield  {journal} {\bibinfo  {journal}
  {Phys. Rev.}\ }\textbf {\bibinfo {volume} {133}},\ \bibinfo {pages} {A895}
  (\bibinfo {year} {1964})}\BibitemShut {NoStop}%
\bibitem [{\citenamefont {Freasier}\ and\ \citenamefont
  {Isbister}(1981)}]{FI81}%
  \BibitemOpen
  \bibfield  {author} {\bibinfo {author} {\bibfnamefont {C.}~\bibnamefont
  {Freasier}}\ and\ \bibinfo {author} {\bibfnamefont {D.~J.}\ \bibnamefont
  {Isbister}},\ }\href@noop {} {\bibfield  {journal} {\bibinfo  {journal} {Mol.
  Phys.}\ }\textbf {\bibinfo {volume} {42}},\ \bibinfo {pages} {927} (\bibinfo
  {year} {1981})}\BibitemShut {NoStop}%
\bibitem [{\citenamefont {Leutheusser}(1984)}]{L84}%
  \BibitemOpen
  \bibfield  {author} {\bibinfo {author} {\bibfnamefont {E.}~\bibnamefont
  {Leutheusser}},\ }\href@noop {} {\bibfield  {journal} {\bibinfo  {journal}
  {Physica A}\ }\textbf {\bibinfo {volume} {127}},\ \bibinfo {pages} {667}
  (\bibinfo {year} {1984})}\BibitemShut {NoStop}%
\bibitem [{\citenamefont {Rohrmann}\ and\ \citenamefont {Santos}(2007)}]{RS07}%
  \BibitemOpen
  \bibfield  {author} {\bibinfo {author} {\bibfnamefont {R.~D.}\ \bibnamefont
  {Rohrmann}}\ and\ \bibinfo {author} {\bibfnamefont {A.}~\bibnamefont
  {Santos}},\ }\href@noop {} {\bibfield  {journal} {\bibinfo  {journal} {Phys.
  Rev. E}\ }\textbf {\bibinfo {volume} {76}},\ \bibinfo {pages} {{051}{202}}
  (\bibinfo {year} {2007})}\BibitemShut {NoStop}%
\bibitem [{\citenamefont {Yuste}\ and\ \citenamefont
  {Santos}(1993{\natexlab{b}})}]{YS93a}%
  \BibitemOpen
  \bibfield  {author} {\bibinfo {author} {\bibfnamefont {S.~B.}\ \bibnamefont
  {Yuste}}\ and\ \bibinfo {author} {\bibfnamefont {A.}~\bibnamefont {Santos}},\
  }\href@noop {} {\bibfield  {journal} {\bibinfo  {journal} {J. Stat. Phys.}\
  }\textbf {\bibinfo {volume} {72}},\ \bibinfo {pages} {703} (\bibinfo {year}
  {1993}{\natexlab{b}})}\BibitemShut {NoStop}%
\bibitem [{\citenamefont {Santos}(2012)}]{LNP_book_note_13_08}%
  \BibitemOpen
  \bibfield  {author} {\bibinfo {author} {\bibfnamefont {A.}~\bibnamefont
  {Santos}},\ }\href@noop {} {} (\bibinfo {year} {2012}),\ \bibinfo {note}
  {``Radial Distribution Function for Sticky Hard Rods'', Wolfram
  Demonstrations Project,
  \url{http://demonstrations.wolfram.com/RadialDistributionFunctionForStickyHardRods/}}\BibitemShut
  {NoStop}%
\bibitem [{\citenamefont {Baxter}(1968)}]{B68}%
  \BibitemOpen
  \bibfield  {author} {\bibinfo {author} {\bibfnamefont {R.~J.}\ \bibnamefont
  {Baxter}},\ }\href@noop {} {\bibfield  {journal} {\bibinfo  {journal} {J.
  Chem. Phys.}\ }\textbf {\bibinfo {volume} {49}},\ \bibinfo {pages} {2770}
  (\bibinfo {year} {1968})}\BibitemShut {NoStop}%
\bibitem [{\citenamefont {Barboy}(1974)}]{B74}%
  \BibitemOpen
  \bibfield  {author} {\bibinfo {author} {\bibfnamefont {B.}~\bibnamefont
  {Barboy}},\ }\href@noop {} {\bibfield  {journal} {\bibinfo  {journal} {J.
  Chem. Phys.}\ }\textbf {\bibinfo {volume} {61}},\ \bibinfo {pages} {3194}
  (\bibinfo {year} {1974})}\BibitemShut {NoStop}%
\bibitem [{\citenamefont {Perry}\ and\ \citenamefont {Throop}(1972)}]{PT72}%
  \BibitemOpen
  \bibfield  {author} {\bibinfo {author} {\bibfnamefont {P.}~\bibnamefont
  {Perry}}\ and\ \bibinfo {author} {\bibfnamefont {G.~J.}\ \bibnamefont
  {Throop}},\ }\href@noop {} {\bibfield  {journal} {\bibinfo  {journal} {J.
  Chem. Phys.}\ }\textbf {\bibinfo {volume} {57}},\ \bibinfo {pages} {1827}
  (\bibinfo {year} {1972})},\ \bibinfo {note} {{N}otice that the numerical
  values in the two last columns of Table I of this reference are not
  correct.}\BibitemShut {Stop}%
\bibitem [{\citenamefont {Tago}\ and\ \citenamefont {Smith}(1977)}]{TS77}%
  \BibitemOpen
  \bibfield  {author} {\bibinfo {author} {\bibfnamefont {Y.}~\bibnamefont
  {Tago}}\ and\ \bibinfo {author} {\bibfnamefont {W.~R.}\ \bibnamefont
  {Smith}},\ }\href@noop {} {\bibfield  {journal} {\bibinfo  {journal} {Can. J.
  Phys.}\ }\textbf {\bibinfo {volume} {55}},\ \bibinfo {pages} {761} (\bibinfo
  {year} {1977})}\BibitemShut {NoStop}%
\end{thebibliography}%
\end{document}